\documentclass[paper]{article}
\usepackage{authblk}
\usepackage[english]{babel}
\usepackage[normalem]{ulem}

\usepackage[usenames,dvipsnames]{xcolor}
\usepackage{amsmath,amsfonts, slashed, amssymb, wrapfig,mathtools}
\usepackage{graphicx}
\usepackage{caption}
\usepackage{subcaption}
\usepackage{verbatim}
\usepackage{cite}

\definecolor{orcidlogocol}{HTML}{A6CE39}
\definecolor{green2}{RGB}{63,142,87}
\definecolor{brown2}{RGB}{136,33,17}
\usepackage{hyperref}
\hypersetup{colorlinks=true, citecolor=orange, urlcolor=blue, linkcolor=magenta}

\begin{document}
\title{The role of bot squads in the political propaganda\\ on Twitter}
\author[1,2,3,4]{\href{https://orcid.org/0000-0001-9377-3616}{Guido Caldarelli}} 
\author[1]{\href{https://orcid.org/0000-0003-4691-7570}{Rocco De Nicola}}
\author[1]{\href{https://orcid.org/0000-0003-1362-7489}{Fabio Del Vigna}} 
\author[1,5,*]{\href{https://orcid.org/0000-0003-0591-877X}{Marinella Petrocchi}}
\author[1]{\href{https://orcid.org/0000-0003-0812-5927}{Fabio Saracco}}

\affil[1]{IMT Scuola Alti Studi Lucca, Piazza S. Francesco 19, 55100 Lucca, Italy \href{mailto:name.surname@imtlucca.it}{name.surname@imtlucca.it}}
\affil[2]{European Centre for Living Technology, Universit\`a di Venezia "Ca' Foscari", S. Marco 2940, 30124 Venice,Italy}
\affil[3]{Catchy srl, Talent Garden Poste Italiane Via Giuseppe Andreoli 9, 00195 Rome, Italy}
\affil[4]{Istituto dei Sistemi Complessi CNR, Dip. Fisica, Universit\`a Sapienza, P.le Aldo Moro 2, 00185 Roma, Italy}
\affil[5]{Istituto di Informatica e Telematica, CNR, Pisa, Italy
\href{mailto:marinella.petrocchi@iit.cnr.it}{marinella.petrocchi@iit.cnr.it}}

\maketitle      
\begin{abstract}
Social Media are nowadays the  privileged channel for information spreading and news checking. Unexpectedly for most of the users, automated accounts, also known as social bots, contribute more and more to this process of news spreading. 
Using Twitter as a benchmark, we consider the traffic exchanged, over one month of observation, on a specific topic, namely the migration flux from Northern Africa to Italy. We measure the significant traffic of tweets only, by implementing an entropy-based null model that discounts the activity of users and the virality of tweets. Results show that social bots play a central role in the exchange of significant content. Indeed, not only the strongest hubs have a number of bots among their followers higher than expected, but furthermore a group of them, that can be assigned to the same political tendency,
share a common set of bots as followers. The retwitting activity of such automated accounts amplifies the presence on the platform of the hubs' messages. 
\end{abstract}
\section{Introduction}
According to Global Digital Report, in 2018, ``more than 3 billion people around the world now use social media each month"\footnote{\url{https://digitalreport.wearesocial.com}}. 
Even traditional newspapers and news agencies moved to social networks, to cope with this societal change.

Since a decade microblogging platforms, like Twitter, have become prominent sources of information \cite{Kwak2010}, catching breaking news and anticipating more traditional media like radio and television \cite{Hu2012}.

Helped by the simple activity consisting of creating a text of 140 (now 280) characters, on Twitter we assist to the proliferation of social accounts governed - completely or in part - by pieces of software that automatically create, share, and like contents on the platform. 
Such software, also known as {\em social bots} - or simply {\em bots} - can be programmed to automatically post information about news of any kind and even to provide help during emergencies.
As amplifiers of messages, bots can simply be considered as a mere technological instrument.
Unfortunately, the online ecosystem is constantly threatened by 
malicious automated accounts, recently deemed responsible for tampering  with online discussions about major political election in western countries, including the 2016 US presidential elections, and the UK Brexit referendum~\cite{gangware2019weapons,Bovet2019,bastos2017brexit,ferrara2015manipulation}. 
Recent work demonstrates that automated accounts are particularly  efficient in spreading low credibility content and amplifying their visibility~\cite{Shao2018}. They also target influential people, bombarding them with hateful contents~\cite{Stella2018}, and they even interact with users according to their political opinion\cite{grinberg2019political}. Bots' actions do not spare financial markets: as much as 71\% of the authors of suspicious  tweets about US stocks have been classified as bots by a state-of-the-art spambot 
detection algorithm~\cite{CresciStock19}.

 Estimates conducted on Twitter report that, on average, social bots account for 9\% to 15\% of total active platform users~\cite{Varol2017}. This notable percentage is highly due to the crucial issue that automated accounts {\it evolve} over time: in a large-scale experiment, it has been proved  that neither Twitter admins, nor tech-savvy social media users, nor cutting-edge applications were able to tell apart evolving bots and legitimate users~\cite{cresci2017paradigm}.
 
 Academicians make their best efforts to fight the never ending plague of malicious bots populating social networks. 
 The literature offers a plethora of successful approaches, based, e.g., on profile-~\cite{cresci2015fame,badri2016uncovering}, network-~\cite{yuan2017spectrum,WangGF17,liu2017holoscope}, and posting-characteristics \cite{Giatsoglou2015,chavoshi2016debot,cresci2018social} of the accounts. In particular, the supervised approach proposed in~\cite{cresci2015fame} tested a series of rules and features from both the grey literature and the official publications on a reference dataset of genuine and fake accounts, leading to the  implementation of a classifier which significantly reduces the cost for data gathering. Propensity to fall into disinformation has been recently measured in US~\cite{ruths2019misinformation,grinberg2019political}, while other work studies the different interactions among users and automated accounts~\cite{Stella2018, Varol2017,Stella2018a}. 
 
During last years, entropy-based null-models for the analysis of complex networks~\cite{Cimini2018,Squartinia} have demonstrated their effectiveness in reconstructing a network from partial information~\cite{Squartini2018}, in detecting early signals of structural changes~\cite{Squartini2013,Saracco2016a} and in assessing the systemic risk of a financial system~\cite{Gualdi2016a,DiGangi2018}. The approach is general and unbiased, being based on the concept of Shannon entropy. In a nutshell, starting from the real network, the method relies on three steps: 1. the definition of an \emph{ensemble} of graphs; 2. the definition of the entropy for this ensemble and its maximization up to some (local or global) constraints~\cite{park2004statistical}; 3. the maximization of the likelihood of the real network~\cite{Garlaschelli2008,squartini2011analytical}.
Recently, such a framework has been  applied to the traffic of messages on Twitter during the 2018 Italian election campaign~\cite{Becatti2019} and it has permitted to infer political standings
directly from data. Moreover, the analysis of the exchanged messages showed a signal of communication between opposite political forces during the election campaign, which anticipated an unexpected post-elections political agreement.

In the present paper, we merge the application of the lightweight classifier for bot detection in~\cite{cresci2015fame} with the analysis of complex networks via entropy-based null-models. 
Once we have cleaned the system from the random noise via the application of the null-model, we study the effects of social bots in retwitting a significant amount of messages on Twitter, without entering in the highly sensitive matter of the veridicity of the messages exchanged.
We apply this analysis to a tweet corpus concerned with the Italian political propaganda about migration in the Mediterranean Sea from Africa to Italy through Lybian ports. We select tweets according to the presence of specific keywords and analyse the network of messages and the related accounts across one month period. We measure that the most effective hubs in the considered 
network have a number of bots followers higher than average. 

In this manuscript, we also find out that a group of successful accounts (hubs), whose owners share similar political views, do share a relatively high number of bots. It appears that the hubs and their bot squads join together, to increase the visibility of the hubs  messages.

\section{Results}
\subsection{Data collection and processing}\label{sec:data}
\begin{center}
\begin{table}[ht]
\begin{center}
\begin{tabular}{l|l}
    \multicolumn{2}{l}{\bf{Keywords}}\\
    \hline \hline
    immigrati&immigrants\\
    migranti&migrants\\
    ong&ngo\\
    scafisti&boat drivers as human smugglers\\
    seawatch& a ngo operating in the Mediterranean Sea\\
    barconi&barges/boats\\
    clandestini&illegal immigrants\\
    guardia costiera libica & Lybian coast guard\\
    naufragio&shipwreck\\
    sbarco&disembarkation\\
    \hline
\end{tabular}
\end{center}
\caption{Keywords used for collecting tweets concerned with Lybia-Italy migrations. Keywords have been searched in Italian, the English translation is on the right. \label{tab:keywords}}
\end{table}
\end{center}
Our study is based on a large corpus of Twitter data, generated by collecting tweets about migrations, and focusing on the case of Lybia-Italy flows.
For the data collection operations, we developed a crawler based on Twitter public Filter API\footnote{\url{https://developer.twitter.com/en/docs/tweets/filter-realtime/api-reference/post-statuses-filter.html}}, which provides real-time tweet delivery, filtered according to specified keywords. We selected a set of keywords compatible with recent chronicles.
Table \ref{tab:keywords} lists the selected keywords. 
The filtering procedure was not case-sensitive. The keywords have been selected because they are commonly used in Italy when talking and writing about 
immigration flows from North Africa to the Italian coasts and about the
dispute about the holder of jurisdiction for handling emergencies, involving European countries and NGOs\footnote{\url{https://en.wikipedia.org/wiki/African\_immigration\_to\_Europe}}. 

We collected 1,082,029 tweets,  posted by 127,275 unique account IDs, over a period of one month (from January 2019, 23rd to February 2019,  22nd).
By relying on the technique introduced in~\cite{cresci2015fame} and recapped in the Methods section, all the accounts have been classified either as genuine or as bots.  This classification led to 117,879 genuine accounts and 9,396 social bots. All  collected tweets were stored in Elasticsearch\footnote{\url{https://www.elastic.co}} for fast and efficient retrieval.

Twitter has the possibility (upon request of the account owner) to give an official certification about the account's `authenticity'. The procedure is mostly adopted by VIPs, official political parties, newspapers, radios and TV channels, in order to reduce interference of fake users. As a result, Twitter users can thus be divided in two sets, the verified and unverified ones\footnote{In the following, we will often adopt the term `validated', not to be confused with `verified'. The former indicates a node that passes the filter of the projection procedure described in the next sections; the latter refers to the aforementioned check by the platform on the identity of an account.}.\\

\subsection{User polarization}
Here, we
consider the bipartite network of verified and unverified users in which links represent an interaction (a tweet or a retweet) among the considered two classes of users.

On Twitter, users are strongly clustered in communities sharing similar ideas (evidences of this, and discussions of its implications, can be found in many papers, see, e.g.,~\cite{DelVicario2017a,Quattrociocchi2014,Zollo2017,Zollo2015}). 
Thus, if we consider two users showing a high number of commonalities among their followers/followees,  they probably have a similar political affiliation. In order to infer directly from data the political orientation of users, after building the (indirect) bipartite network of verified/unverified users, we project the bipartite network on the layer of the verified users (see the Methods section) and consider the number of contacts shared by verified accounts.
The presence of a strong community structure in the bipartite networks of verified/unverified users has already been observed\cite{Becatti2019}; we repeat this analysis here and check results on the layer of verified users, for which we have  reliable information. We confirm that also in the present case the result of~\cite{Becatti2019} holds, with some caveats, as we will see in the following. 

\subsubsection{Political affiliation}
\begin{figure}[ht!]
    \centering
    \includegraphics[width=.7\linewidth]{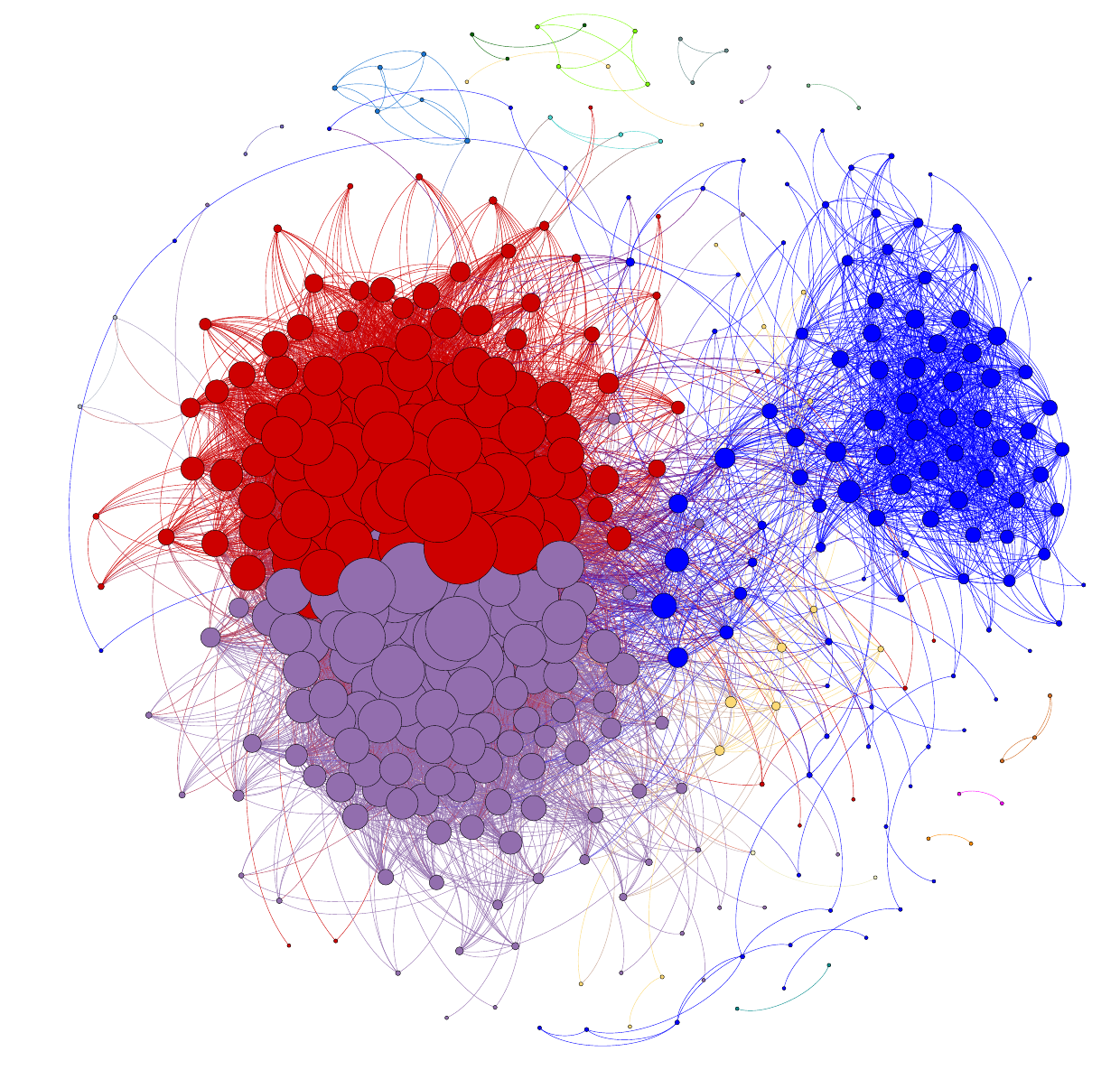}
    \caption{The network resulting from the projection procedure. In \textcolor{blue}{blue}, accounts tied to the Italian government\protect\footnotemark[5], to the right wing\protect\footnotemark[6], to Movimento 5 Stelle\protect\footnotemark[7] and the official account of the newspaper `Il Fatto Quotidiano' with its journalists~\cite{Becatti2019}. In \textcolor{red}{tomato red}, the accounts of the Italian Democratic Party  (PD) and its representatives\protect\footnotemark[8], as well as some representatives of smaller parties on the left of PD\protect\footnotemark[9]. The {\color{violet} purple} group includes several NGOs\protect\footnotemark[10], politicians on the left of PD\protect\footnotemark[11], different online and offline newspapers\protect\footnotemark[12]. In {\color{orange} orange}, some official accounts related to the Catholic Church\protect\footnotemark[13]. Finally, in {\color{teal} turquoise} we found smaller groups related to the Malta government (including the Prime Minister Joseph Muscat and some of his ministers), and in {\color{green2} green} we even found a soccer commentators  community.}
    \label{fig:affiliation_netwk}
\end{figure}
The bipartite network describing the interactions between verified and unverified users involves nearly one half of the unverified users in our dataset. This limited interaction differs from the one in~\cite{Becatti2019}, where almost all nodes were involved in the polarization analysis. Nevertheless, even if here we have to restrict our study to half of the accounts, the system still displays strongly structured communities.
In fact, the network obtained following the projection procedure 
described in the Methods shows a strong community structure, see Figure~\ref{fig:affiliation_netwk}.

\footnotetext[5]{We find in this set the Minister for the Internal Affairs Matteo Salvini, the Minister of Infrastructures and Transports Danilo Toninelli, the Minister of Health Giulia Grillo.}
\footnotetext[6]{The right wing verified accounts  include the right party Fratelli d'Italia and its leader Giorgia Meloni, Forza Italia and some of its representatives as Giovanni Toti, as well as Lega with other politicians than the above mentioned Salvini.}
\footnotetext[7]{For instance,  Carlo Sibilia and Roberta Lombardi can be found here.}
\footnotetext[8]{The candidates to the PD primary elections Nicola Zingaretti, Maurizio Martina and Roberto Giachetti, its previous leader Matteo Renzi and the previous Prime Minister Paolo Gentiloni contribute to this community.}
\footnotetext[9]{For example, Pierluigi Bersani, Laura Boldrini and Marco Furfaro are in this group.}
\footnotetext[10]{In the NGOs set we identified the Italian chapters of ActionAid, Medecins Sans Frontieres, Unicef,  Amnesty International, OpenArms.}
\footnotetext[11]{Giuseppe Civati, Nichi Vendola, Luigi De Magistris and Enrico Rossi are part of this set.}
\footnotetext[12]{A variety of accounts can be found in this set: Repubblica, il Post, Internazionale, la Valigia Blu, Rolling Stone Italia and their journalists, many public figures as the film directors Gabriele Muccino and Ferzan Ozpetek, the actors Alessandro Gassmann and Pif, the singers Fiorella Mannoia, Nina Zilli, Frankie Hi-NRG}
\footnotetext[13]{The newspaper `l'Osservatore Romano' and the website `Vatican News' and `Vatican Radio' are part of this group.}

To quantify the presence of clusters, we decide to use one of the most effective and used community detection algorithms, that is the Louvain  algorithm~\cite{Blondel2008}. To avoid the problems related to the scale of applicability of such algorithm~\cite{Fortunato2010}, we apply the algorithm after reshuffling the order of nodes for $N$ times ($N$ being the number of nodes in the network). The corrected algorithm returns 3 big communities, see Figure~\ref{fig:affiliation_netwk}: the description of the community by membership can be found in the caption.

With respect to the previous study on the election campaign~\cite{Becatti2019}, here we find several differences. While the M5S group and their supporters were distinguishable from the right wings during the election campaign, in the propaganda regarding the Mediterranean migration it is not. Similarly, the left wing representatives, both inside and outside the Democratic Party, are much closer than during the election campaign. Smaller communities regards Malta prime Minister Joseph Muscat and part of his ministers involved in the discussion for the aid of migrants and castaways.

\subsubsection{Polarization of unverified accounts}
\begin{figure}[ht!]
    \begin{minipage}{.4\textwidth}
    \includegraphics[width=\linewidth]{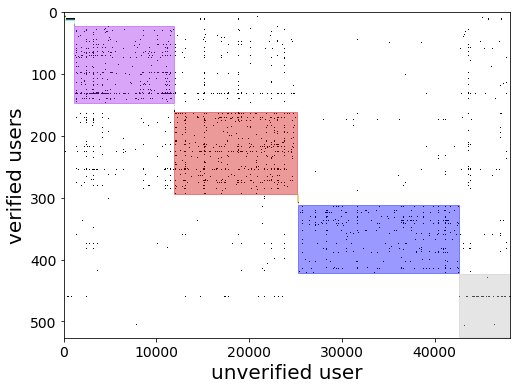}
    \end{minipage}
    \hfill
    \begin{minipage}{.4\textwidth}
    \includegraphics[width=1.2\linewidth]{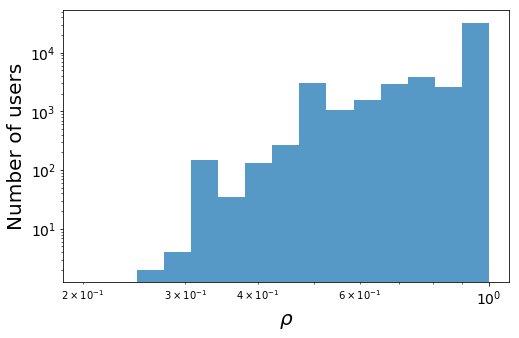}
    \end{minipage}
    \hspace*{\fill}
    \caption{Left panel: biadjacency matrix, describing the interactions between verified and unverified users. Nodes have been rearranged in order to highlight the community structure and colored according to the communities of Figure~\ref{fig:affiliation_netwk}. In gray, users with lower value of polarization or not projected by the validated projection (see Methods and Supplementary Materials). Right panel: polarization index distribution among unverified users. The plot is in log-log scale, i.e., nodes with a polarization higher than 0.9 are more than 10 times the one of those with polarization between 0.8 and 0.9.}
    \label{fig:polarization}
\end{figure}

Verified accounts of politicians can be easily associated to membership of a  political party, for unverified users we assign membership by considering their interactions with the community including the verified ones. 
We use the polarization index $\rho_i$ as defined in~\cite{Bessi2016,Bessi2016a}:
\begin{equation}
    \rho_i=\dfrac{\max_{c\in\mathcal{C}}k_i^c}{k_i},
\end{equation}
where $k_i$ is the degree of node $i$, $k_i^c$ is the number of links towards the community $c$ and $\mathcal{C}$ is the set of communities. The distribution of $\rho_i$ is extremely peaked on values close to 1, see Figure~\ref{fig:polarization}, right panel. Given such a strong polarization, we can safely assign unverified users the polarization of the community they mostly interact with.

We also find a smaller amount (with respect to the size of verified-unverified network) of unverified nodes whose polarization is not strong enough to be uniquely assigned to a specific cluster: they are part of the gray group in Figure~\ref{fig:polarization}. 

As noted above, almost one half of the unverified users do not enter into this polarization procedure, not interacting in the whole period with a single verified user. This may be due to several reasons. Differently from~\cite{Becatti2019}, where a corpus of tweets exchanged during the election campaign was analyzed, here we focus on a set of tweets concerned with a specific topic of the political propaganda. 
We conjecture that the amount of unverified accounts interacting with the verified ones in the former case was much higher because it was of interest of the verified accounts (mostly, candidates in the elections) to involve `standard' users. 

In order to know more about those unverified users not directly interacting with the verified ones, we introduced a {\em contagion of polarization}. Unverified users do indeed interact with each other and therefore, if the majority of their (unverified) interacting partners is polarized, then we can use this fact  to infer the polarization of the users.
We can iteratively assign to unverified users the prevailing polarization of the other accounts they interact with and stop when there is no possibility to assign a polarization anymore, i.e., if there is no clear agreement among the neighbours of the considered node. In this way, after 10 rounds of such a procedure, we are able to increase the fraction of the users for which we determine a clear political membership by 27\%. Even if this percentage may seem quite small, we will see that the aforementioned contagion process is more effective on the set of validated accounts, presented in the following sections. In this latter case, the increase of polarized users is almost 58\%; more details can be found in the Supplementary Materials. 
In the plots in the following sections, the community colors are those obtained by relying on polarization by contagion.

\begin{figure}[ht!]
    \centering
    \begin{minipage}{.45\textwidth}
    \centering
    \includegraphics[width=\linewidth]{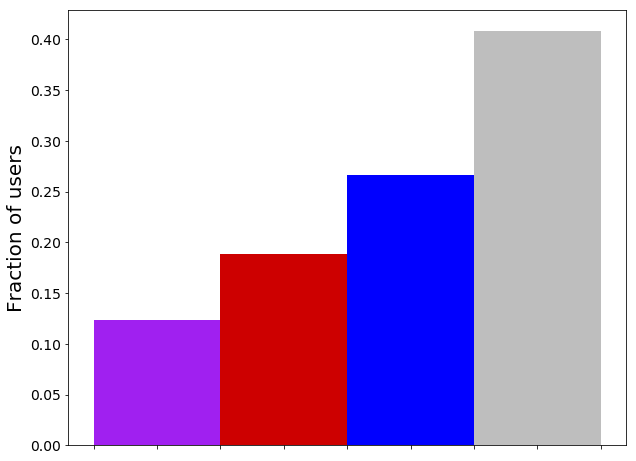}
    \end{minipage}
    \hfill
    \begin{minipage}{.45\textwidth}
    \centering
    \includegraphics[width=1\linewidth]{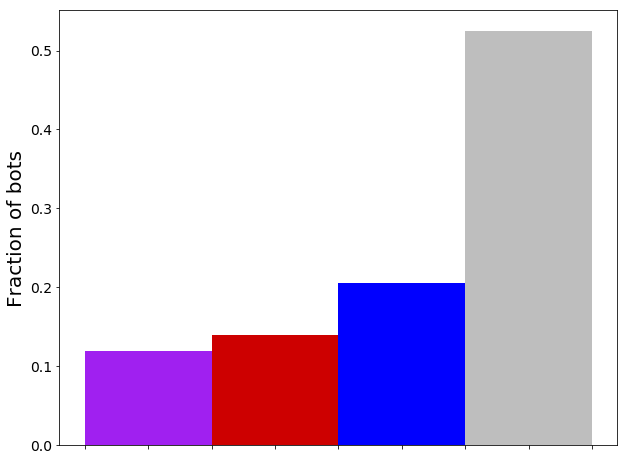}
    \end{minipage}
    \caption{Left panel: the histogram of the 4 main communities for polarization of all users after the iterative polarization. Nodes have been  colored according to the partition in Figure~\ref{fig:affiliation_netwk}. In gray, users which cannot be assigned uniquely to one community. Right panel: the same histogram for bots, as recognized by the bot detector. Percentages of the different non-gray groups are lower than the analogous for all users, but for the purple community. The (relative) extra density of bots in the purple community may be due to the high presence of automated news spreaders, used for increasing the visibility of the articles of the newspapers accounts.}
    \label{fig:humans_vs_bots}
\end{figure}

Interestingly enough, assigning a polarization to bots is much harder than to average users: Figure~\ref{fig:humans_vs_bots} illustrates the density of all users (left panel) and bots (right panel) in the 4 biggest communities and shows that it is much more difficult to assign a polarization to bots. Neglecting the contribution of the gray bars, the relative fractions of the different communities are more or less similar, but for a slight increase of the abundance of `purple' bots.

\subsection{The backbone of the content exchange on Twitter}
In the analysis of a complex system, one of the main issues is to skim the relevant information from the noise. Of course, the definition itself of noise depends on the system. In the present study, via the use of an entropy-based null model~\cite{Cimini2018, Squartinia, park2004statistical,Garlaschelli2008, squartini2011analytical}, we filter the total exchange of content in our dataset, discounting the information regarding the activity of users and the virality of messages. Literally, we start by considering the directed bipartite network of users (on one layer) and tweets (on the other layer): arrows are directed from the user to the tweet in the case the user is the author of the message, while in the opposite direction indicates that the user retweets the given message. We then construct the Bipartite Directed Configuration Model~\cite{DeJeude2018}, i.e., the extension of the standard framework of entropy-based configuration model~\cite{Cimini2018, Squartinia, park2004statistical,Garlaschelli2008, squartini2011analytical} for bipartite directed networks. In the present case, the constraints are represented by the node activity, i.e., the number of original tweets posted by every user, the number of retweets of every message and the number of retweets of every user. Thus, by comparing the real system with the null-model, we can highlight all the contributions that {\it cannot be explained only by the fixed constraints}. More details can be found in the Methods section and in the Supplementary Materials.

As in the case of the polarization above, we can thus filter the flow of information on the network from the random noise due to the activity of users and to the virality of tweets. The filtering returns a directed network in which the arrows go from the authors to the retwetters and it reduces the number of nodes to 14,883 users and of links to 34,302.

\begin{figure}[ht!]
    \centering
    \includegraphics[width=.8\linewidth]{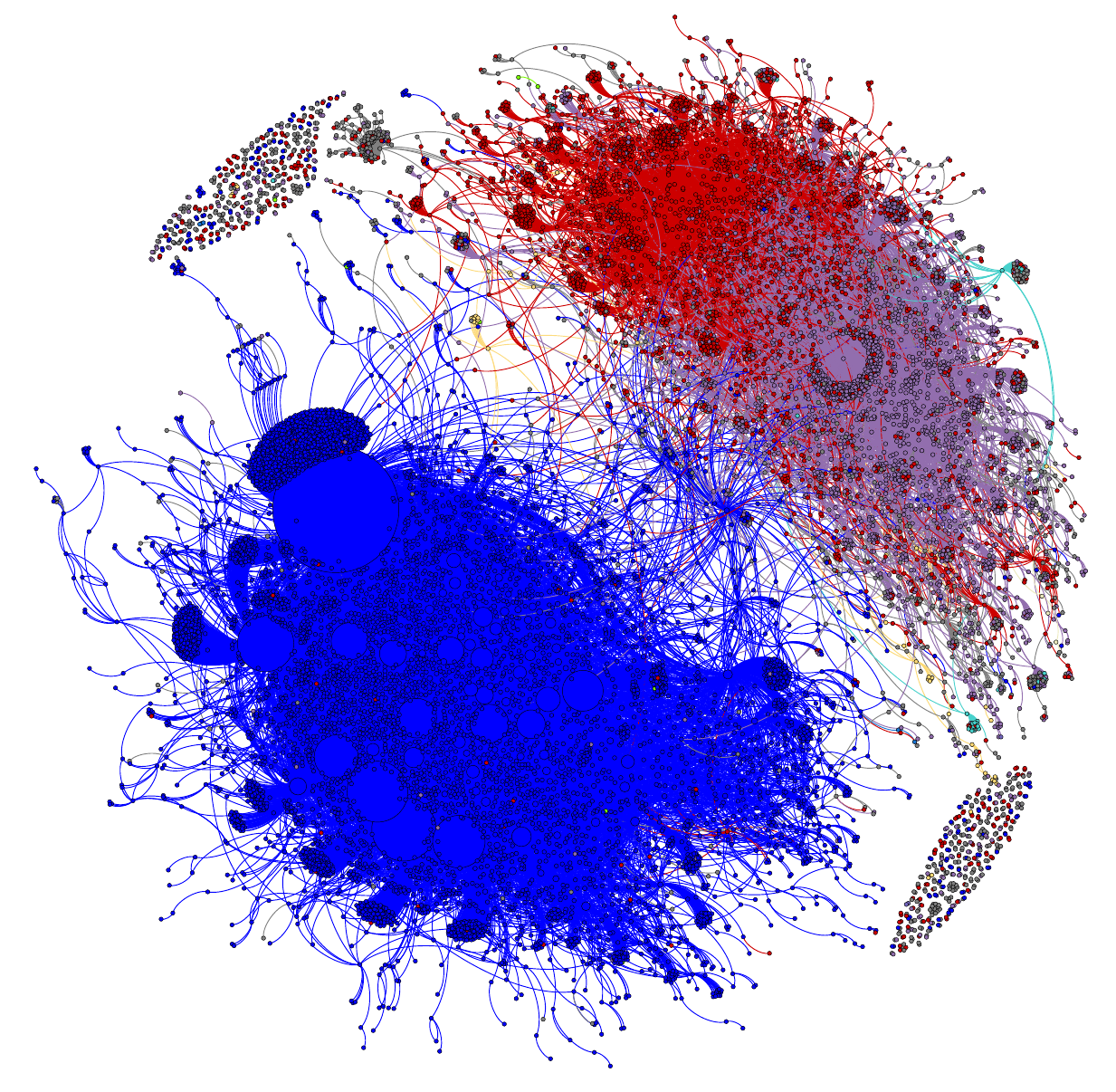}
    \caption{The directed validated network. Nodes have been  colored according to the partition in Figure~\ref{fig:affiliation_netwk}. The dimension of each node is proportional to its hub score: the biggest node (in blue) is the account of Matteo Salvini.}
    \label{fig:validated_netwk}
\end{figure}
On top of this, we analyze the presence of automated accounts, by using the bot detection method described in the Methods. The incidence of bots on the number of nodes is about 2.5\%, against almost the 7\% of nodes in the original network. The number of loops, i.e., users that retweet their own posts, (significantly with respect to their activity), is circa the 1.2\% over the total amount of links, thus relatively high. This effect reverberates also on  the number of validated nodes, that significantly retweet  themselves (little less than the 3\%). For the subsequent analysis, we discard the contribution of loops, since we are interested in analysing the source of the shared contents on Twitter.

\subsubsection{Hubs and bots}
As mentioned in the previous section, the validated links go from the authors to the retwitters. In this sense, the effectiveness of an author can be derived by its ability to reach a high number of most relevant nodes: this principle is finely implemented in the Hubs-Authorities algorithm~\cite{Kleinberg1999, caldarelli2010scale-free,Newman2010}. Authorities, in this analysis, are sort of the sink of the content exchange. Since, in the present system, the strongest authorities are represented by unverified accounts, it is quite difficult to interpret the results. In the following, we focus on hubs, because they represent the driving force of the discussion and are relatively popular users; thus even if they are not verified by Twitter, we often have reliable information about their accounts.

\begin{figure}[ht!]
    \centering
    \begin{minipage}{.45\textwidth}
    \centering
    \includegraphics[width=\linewidth]{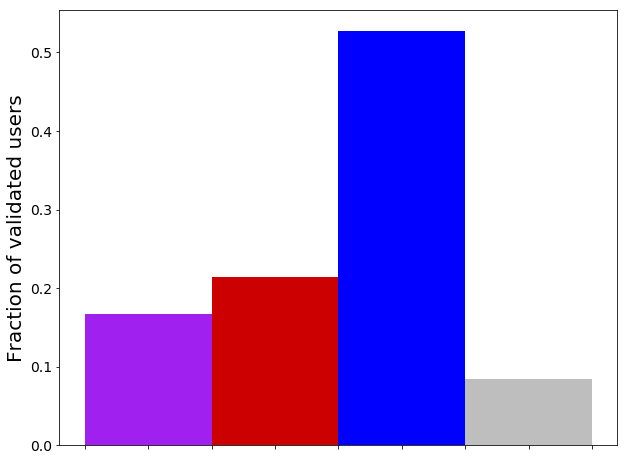}
    \end{minipage}
    \hfill
    \begin{minipage}{.45\textwidth}
    \centering
    \includegraphics[width=1\linewidth]{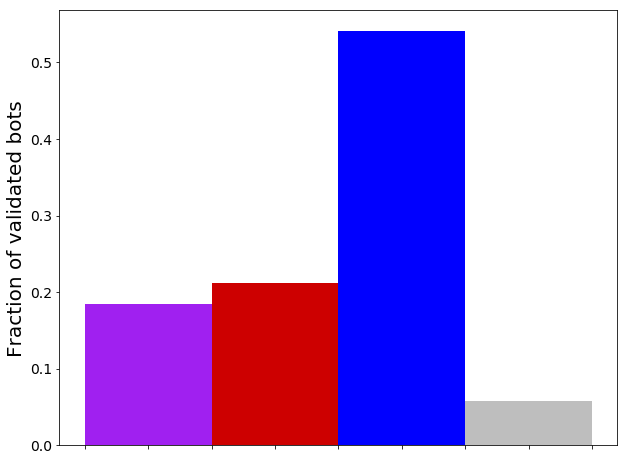}
    \end{minipage}
    \caption{Left panel: the histogram of the 4 main communities for polarization of all users in the validated network after the iterative polarization. Nodes have been  colored according to the partition in Figure~\ref{fig:affiliation_netwk}. Right panel: the same histogram for bots in the validated network. By comparing the present histograms with those of Figure~\ref{fig:humans_vs_bots},  it can be seen that, in this case, the number of validated accounts for which the polarization procedure is not able to produce a unique output is much smaller. Indeed, most of the unpolarized accounts do not contribute significantly to contents exchange in the validated network.}
    \label{fig:humans_vs_bots_validate}
\end{figure}

\begin{table}[t]
\centering
\begin{tabular}{c|cccc}
Screen name & $\text{Hub}_\text{score}$ & $k^\text{out}_i$ & $\dfrac{|\text{bot}_i|}{k^\text{out}_i}$ & $\dfrac{|\text{bot}_i|}{k^\text{out}_i}\Big/\dfrac{|\text{bot}|}{N_\text{validated users}}$\\
\hline
\hline
`matteosalvinimi'& 1.000 & 3473 & 0.023 & 1.058\\
`hub\_1' & 0.490 & 1270 & 0.003 & 0.141\\
`hub\_2' & 0.465 & 1199 & 0.004 & 0.187\\
`GiorgiaMeloni' & 0.427 & 1303 & 0.032 & 1.444\\
`hub\_4' & 0.395 & 1040 & 0.005 & 0.215\\
`hub\_5' & 0.326 & 809 & 0.011 & 0.498\\
`hub\_6' & 0.300 & 775 & 0.009 & 0.404\\
`hub\_7' &0.290 & 574 & 0.002 & 0.078\\
`hub\_8' & 0.282 & 583 & 0.0 & 0.0\\
`hub\_9'& 0.271 & 646 & 0.003 & 0.139\\
`hub\_10' & 0.200 & 395 & 0.005 & 0.227\\
`hub\_11' & 0.189 & 368 & 0.0 & 0.0\\
`hub\_12' & 0.186 & 401 & 0.005 & 0.224\\
`hub\_13' & 0.166 & 341 & 0.009 & 0.394\\
`hub\_14' & 0.152 & 268 & 0.0 & 0.0 \\
`hub\_15' & 0.133 & 245 & 0.012 & 0.549\\
`hub\_16' & 0.128 & 222 & 0.0 & 0.0\\
`hub\_17' & 0.126 & 299 & 0.013 & 0.600\\
`hub\_18' & 0.112 & 190 & 0.0 & 0.0\\
`hub\_19' & 0.106 & 279 & 0.011 & 0.482\\
\hline
\end{tabular}
\caption{Screen names of the hubs in the validated network, their hub score, their out-degree $k^\text{out}$, the fraction of bots in their out-neighbours (indicated as $\frac{|\text{bot}_i|}{k^\text{out}_i}$ and the ratio between this value and the average over the entire network (indicated as $\frac{|\text{bot}_i|}{k^\text{out}_i}\Big/\frac{|\text{bot}|}{N_\text{validated users}}$). For the sake of privacy, the screen names of the unverified accounts have been anonymized.}
\label{table:hubs_description}
\end{table}

Table~\ref{table:hubs_description} shows the values for the top 20 nodes, in term of hub scores. The first  account is the one of the Italian Minister of Internal Affairs Mr. Matteo Salvini. 
The second and the third ones refer to two journalists of 
a news website supported by Casa Pound, 
a neo-fascism Italian party. The fourth is the account of Ms. Giorgia Meloni, leader of the right political party Fratelli d'Italia, former ally during 2018 Italian political elections of the Lega, the political party of Mr. Salvini. The political vision of the two leaders is quite close regarding the theme of ruling the Mediterranean migration. The two accounts following in the rank are respectively  a journalist of `Il Fatto Quotidiano' (newspaper supported by M5S) and an unverified user with opinions in line with the ones of the two above mentioned politicians. All the accounts in Table~\ref{table:hubs_description} belong to the blue community. 
The first account with a different membership (`TgLa7', a popular newscast by a private TV channel, in the purple community in our analysis) ranks $176^\text{th}$ with respect to the hub score. It is  striking the case of the Italian chapter of 
a NGO assisting migrants in the Mediterranean Sea: while it has the fifth highest value of out-degree ($k_\text{out}=1104$), it has an extra low hub score ($4\times10^{-4}$), ranking $452^\text{nd}$. It is even more impressive, considering that, in several occasions, the Italian government (in the figure of the Minister of Internal Affair, mostly) and the NGO have been opponents regarding the disembarkation of the migrants rescued during the NGO activities. 

\begin{figure}[ht!]
    \centering
    \includegraphics[width=\linewidth]{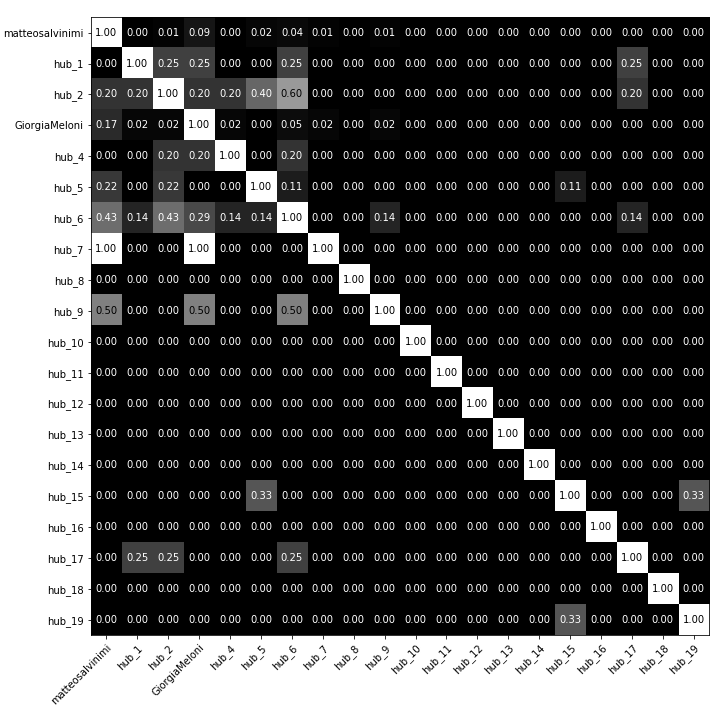}
    \caption{The relative overlap matrix among list of bots following the different hubs.
    Using the formalism of the previous Table~\ref{table:hubs_description}, the generic matrix entry represents $\frac{|\text{bot}_i \cap\text{bot}_j|}{|\text{bot}_i|}$. There are 12 accounts sharing a relatively high number of bots.}
    \label{fig:overlap_matrix}
\end{figure}

\begin{figure}[ht!]
    \centering
    \includegraphics[width=\linewidth]{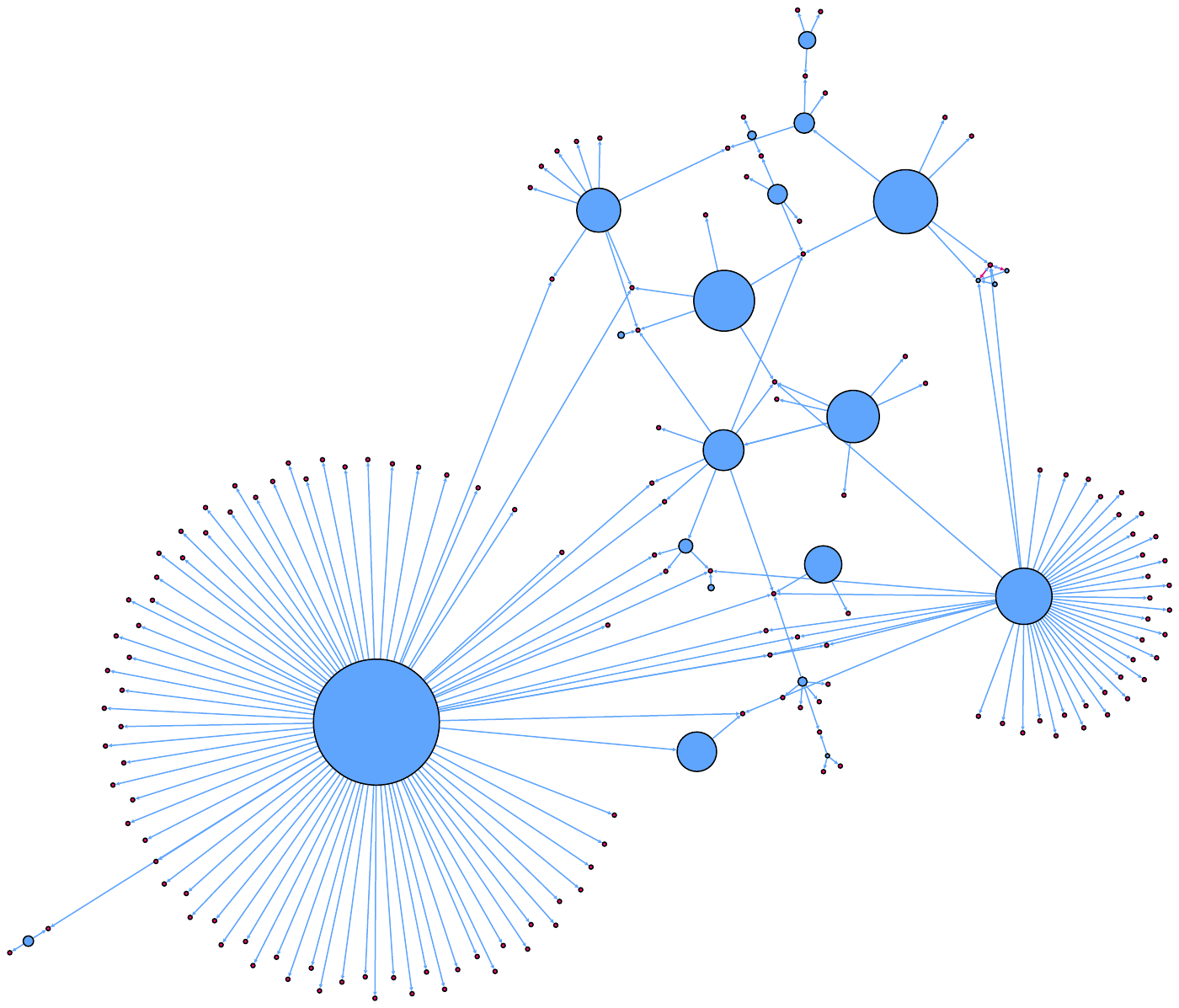}
    \caption{The subgraph of the greatest group of users sharing bots. The subgraph includes human accounts (in dark blue) and all the bots following them (in light red). The dimension of the nodes is proportional to the hub scores, but normalised on the subgraph. The biggest node represents the account of Mr. Salvini. In the picture, there are 22 bots shared by 22 humans. Among the latter, 
    9 accounts are in the list of top 10 hubs. The subgraph contains 172 nodes. Notably, accounts belongs almost exclusively to the blue community.}
    \label{fig:bot_corps_0}
\end{figure}

\begin{figure}[ht!]
    \centering
    \includegraphics[width=\linewidth]{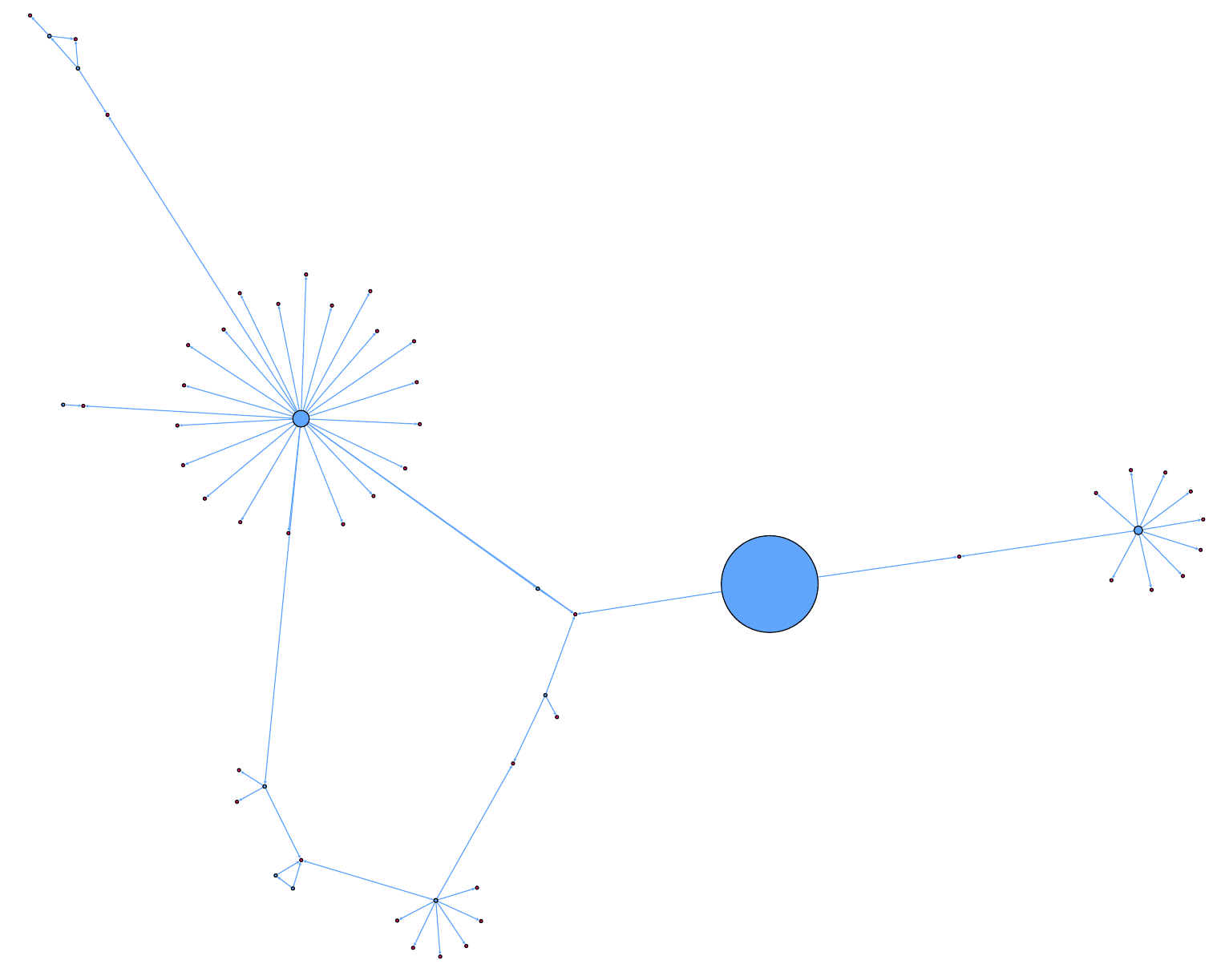}
    \caption{
  The subgraph of the second greatest group of users sharing bots. Subgraphs include human accounts (in violet) and all the bots following them (in light red). The dimensions of the bots are proportional to the hub scores, but normalised on the subgraph. Thus, the dimension of the greatest node in Figure \ref{fig:bot_corps_0} should be 362 times bigger than the greatest one here. The impact on the validated network of the nodes in this graph is much more limited: the strongest hub, i.e., the biggest node in the plot, is the official account of `TgLa7', a newscast ranking $176^\text{th}$ in the hub scores. Considering even the non shared bots, the subgraph contains 58 nodes. Notably, accounts belongs almost exclusively to the purple community.}
    \label{fig:bot_corps_1}
\end{figure}

Remarkably, we observe a non zero overlap among the bots in the list of the validated followers of human users. To the best of our knowledge, this is the first time that such a phenomenon is detected. The presence of bot squads, retweetting the messages of two or more strong hubs, increases the visibility of their tweets.

We detect two main groups of such accounts, the other being composed by a maximum of 2 common bots. The first one includes 22 genuine accounts (9 of which are in the top 10 hubs), sharing 22 bots. In this set, some users share a relatively high fraction of bots: there is the case of a right wing account sharing all his automated followers with both Meloni and Salvini, see Figure~\ref{fig:overlap_matrix}. In Figures~\ref{fig:bot_corps_0} and \ref{fig:bot_corps_1}, we represent two subgraphs of the original validated network of Figure~\ref{fig:validated_netwk}. Figure~\ref{fig:bot_corps_0} shows the first group of genuine accounts sharing bots and all their bot followers. The hub scores, represented as the dimensions of the nodes, are nearly homogeneous among the hubs. This does not happen in the analogous subgraph for the second group (see Figure~\ref{fig:bot_corps_1}): beside the presence of a strong hub, the hub score distribution is much skewer than for the previous group. Moreover, in absolute terms, the hub scores are much smaller than the previous case, since the strongest hub is the aforementioned account of `TgLa7' newscast. More details about the composition and the features of the two groups can be found in the caption of Figures~\ref{fig:bot_corps_0} and \ref{fig:bot_corps_1}.

\section{Discussion}
The 2018 Eurobarometer report on news consumption presents  a clear increasing trend of popularity of online sources with respect to traditional ones~\cite{TNSopinionsocial2018}. Albeit this widespread favour, online media are not trusted as their offline counterparts: 
in a survey conducted in autumn 2017, 59\% of respondents said they trusted radio content, while only 20\% said they trusted information available on online social networks.
Even beside the perception of common users, the presence of fake contents has indeed been revealed in several research work, both at level of news {\it per se} and of fake accounts contributing to spreading them, see, e.g.,~\cite{Quattrociocchi2014,cresci2015fame,Ferrara2016rise,Shao2018, cresci2018social}.

Among different platforms, Twitter is one of the most studied, due to the openness of its data through the public APIs. Also, it is strongly used for political propaganda: a recent survey~\cite{AGCOM2017} showed that Italian journalists appear on Twitter much more frequently than common users. Therefore,  Twitter has been used for many analyses of communication in the political propaganda, see, e.g.,~\cite{Becatti2019,Stella2018,Stella2018a,DelVicario2017a,Shao2018, Bovet2018,Bovet2019,Bekafigo2013,Borondo2012,DelVicario2018,DiGrazia2013,Gonzalez-Bailon2011,Gonzalez-Bailon2013,Ferrara2017}. Obviously, a major issue when performing such kind of analyses is the reliability of the results, which is closely connected to the reliability of the users in the game: in such sense, a rich stream of research is devoted to find powerful means for detecting automated accounts - even anticipating their future evolution~\cite{cresci2019websci} - and their interactions with human-operated accounts~\cite{Stella2018,Stella2018a,Balestrucci2019}.

Remarkably, all the previous analyses rarely tackle the effect of random noise, which is indeed of utmost importance when studying complex systems. In~\cite{Jaynes1957}, Jaynes showed  how Statistical Physics could be derived from Information Theory from an entropy maximization principle. Following Jaynes work, in recent years the same approach has been extended to complex networks~\cite{Cimini2018, Squartinia, park2004statistical,Garlaschelli2008, squartini2011analytical},  to provide an unbiased benchmark for the analysis, by filtering out random noise. Such a framework proved to be 
extremely ductile and adaptable to the description of different phenomena~\cite{Fronczak2013a,Mastrandrea2014,Saracco2015a,Becatti2018,DeJeude2018}. In the present study, we merge the application of bot detection techniques with the use of an entropy-based null-model for the analysis of the content exchange on Twitter in the Italian discussions about regulating the migration flux from Northern Africa. The corpus we
analyzed resulted to be extremely informative in highlighting some otherwise hidden features of the dissemination of information in those discussions.

First, in order to get the political affiliation of users, we focused on the bipartite network in which the two layers represent verified and unverified users, respectively,  and the (undirected) links label the interactions between the two classes. The main idea is to infer the 
inclination of users towards a political point of view
from (a proxy of) their contacts: if two users share a great amount of followers and followees, they probably have a similar political polarization. Following the strategy of~\cite{Saracco2016a}, we use the above mentioned entropy-based framework to project the bipartite network on the layer of verified users, whose account information is reliable. 
Verified users have been clustered into 3 main groups, see Figure~\ref{fig:affiliation_netwk}: one group includes government representatives, the right wing and the Movimento 5 Stelle party; a second group includes the Italian Democratic party; a third one includes NGOs, online and offline media, journalists and some VIPs.
Confirming results presented in other studies~\cite{Quattrociocchi2014,Schmidt2018,DelVicario2017a,Nikolov2015}, the polarization of unverified users is particularly strong: they interact quite exclusively with accounts of a single community, see Figure~\ref{fig:polarization}. 
Differently than in other studies~\cite{Becatti2019}, the interaction of unverified users with verified ones is limited, and affects only one half of the total amount of unverified users. This is probably due to the fact that we focus on a discussion that is wider than an election campaign and that could stimulate exchanges between users who do not usually participate in political discussions. Thus, we iteratively assign group memberships to unverified users, based on  the political affiliation of the majority of all their followers and followees. This procedure reduces the number of unpolarized accounts of more than 35\%. Curiously, the ratio of bot accounts that remain unpolarized after the `political contagion' is higher than the analogous for all users, see Figure~\ref{fig:humans_vs_bots}. In any case, in the following, we will see that users, automated or not, taking \emph{effectively} part to the discussion are mostly polarized.

Finally, we extract the non trivial content exchange by adopting the validated projection developed in~\cite{Becatti2019}: this permits to detect the significant flow of messages among users, discounting, at the same time, the virality of messages, the retweetting activity of users and their productivity in writing tweets. Such an approach provides the `backbone' of the content exchange among users on Twitter.


The network represented in Figure~\ref{fig:validated_netwk} is extremely informative for different reasons. 
The validated network contains only 14,883 validated users out of the 127,275 users in the dataset. This  highlights the fact that just a minority of all users effectively contributes  to the online propaganda on the migration flow. Interestingly, we found that the incidence of bots on the validated network is almost one third of the analogous measure on the entire dataset,
signaling that the number of bots whose retweets are non compatible with a random activity is just a small minority. Since the target of a social bot is to increase  audience of the online content of a specific user, such a reduction shows that the number of bots affecting significantly the political discussion is limited.

The set of validated users is much more polarized than the whole set of users, see Figure~\ref{fig:humans_vs_bots_validate}. We have that the  overall fraction of unpolarized accounts represents more than 40\% of all the accounts and more than 50\% of the automated ones, while when considering the validated network, the same ratio is around 10\% for the former and around 5\% for the latter. 
Otherwise stated, the polarized bots pass  the validation process more easily than their unpolarized counterparts and their contribution in spreading messages is more significant.

All the accounts that are mostly effective in delivering their messages (i.e., the Hubs~\cite{Kleinberg1999})  refer to the blue area in Figure~\ref{fig:validated_netwk}, where we can find representatives of the the government in charge and the right wing.  
 The first account referring to a community different from the blue one is the official account of the newscast `TgLa7', at position $176^\text{th}$ in the hub ranking.
 
The contribution of bots to the visibility of the various accounts shows that the fraction of bots
 that significantly retweet the content of two right wing political leaders (Mr. Salvini and Ms. Meloni)
 is greater than the incidence of bots in the validated network.
 Interestingly enough, other hubs show a smaller presence of bots among their followers, even if their hub score is not that different from  the two political leaders.
 
 Finally, we have that some hubs do share their bots: Figure~\ref{fig:overlap_matrix} describes the normalized overlap  between the list of bots of each pair of users in the list of the top 20 hubs. As mentioned before, those accounts are from the right wing political area. To the best of our knowledge, this is the first time that such a behaviour is reported: in analyses tackling the same problem~\cite{Varol2017,Stella2018,Stella2018a}, only star-like sub-graphs were observed, with a big number of bots among the followers of a (presumably) human user. We underline that the considered shared bots are particularly effective, since they are validated by the entropy-based projection. Actually, the group of ``right wing" bots, each supporting more than a human account, is not the only one in the set, but it is the greatest: if we consider the subgraphs of human accounts sharing their bots - see Figures~\ref{fig:bot_corps_0}, \ref{fig:bot_corps_1} -, the former has 172 nodes against 58 of the latter. Moreover the first subgraph is by far more efficient; indeed, in the second one the greatest hub score ranks $176^\text{th}$. 

It is well known that  bots aim at increasing popularity of users by retweetting their messages, see, e.g.,~\cite{cresci2017paradigm}: exactly what is revealed by the entropy-based filtering. The latter turns out to be extremely helpful, since {\it it hits one feature of an automated account that cannot be avoided by programmers}. To the best of our knowledge, the study here presented is the first investigation that merges  bot detection and entropy-based analysis of Twitter traffic.
Moreover, the obtained results are in line with the previous work of~\cite{Shao2018}, where the authors showed how bots massively support the spread of (low credibility) content. At the same time, the present outcome contributes in a different way, being not specifically focused on fake news, whereas~\cite{Shao2018} concentrates on the way fake news become viral. Interestingly enough, among the many studies of the 2016 US presidential election, Grinberg {\it et al.}~\cite{grinberg2019political} analyzed the proliferation of fake news on Twitter and determined both fake news spreaders and exposed users. Remarkably, it was found that fake news was `most concentrated among conservative voters'. 
The role of bots in effectively conveying a message - for the first time here highlighted even in a `shared fashion' - and the spreading of fake news in online discussions of great importance~\cite{grinberg2019political,Shao2018} leads us to a promising future direction of study, which include a deeper  analysis of the exchanged messages, like the extraction of their sentiment and the contained mentions. 



\section{Methods}\label{methods}
\subsection{Bot detection classifier}
To assess the nature of the accounts in the dataset about migration from Northern Africa, we rely on a slightly modified version of the supervised classification model proposed in~\cite{cresci2015fame}.

In that work, a series of machine learning algorithms were originally trained on a baseline dataset of genuine and fake accounts. The latter were bought on three different Twitter accounts online markets, while the former were certified as genuine by tech-savvy social media analysts.

The features and rules tested for the classification were among the most relevant ones proposed by Academia for anomalous Twitter accounts detection. 
Looking at the performances of the various algorithms on the training dataset, the final result was a novel  classifier, 1) general enough to thwart overfitting (i.e., the problem of being too
specialized on the training dataset and unable to generalize the classification to new data), 2) lightweight, thanks to the usage of features that require only information present in the profile of the account and 3) able to correctly classify more than 95\% of the training data.

For the present paper, we reconstruct the model of the classifier in~\cite{cresci2015fame} and we test its performances with J48, the Weka\footnote{\url{https://www.cs.waikato.ac.nz/ml/weka/}} implementation of C4.5 algorithm, on the same training set\footnote{Dataset publicly available at \url{http://mib.projects.iit.cnr.it/dataset.html}}, obtaining the same classification performance results. The used features are listed in Table~\ref{tab:features}.

\begin{center}
\begin{table}[ht]
\begin{center}
\begin{tabular}{l}
    \multicolumn{1}{l}{\bf{Features}}\\
    \hline \hline
    friends count\\
    followers count\\
    tweets count\\
    $\frac{friends}{followers^2}$\\
    account age\\
    following rate (approximated as $\frac{friends}{age}$)\\
    the account's profile has a name\\
    the account's profile has an image\\
    the account's profile has an address\\
    the account's profile has a biography\\
    the account's profile has a URL\\
    the account belongs to a list\\
    $2 \times followers \geq friends$\\
    $100 \times friends \geq followers$\\
    $50 \times friends \geq followers$\\
    \hline
\end{tabular}
\end{center}
\caption{Features adopted for the fake account detector designed in~\cite{cresci2015fame} and here re-constructed\label{tab:features}.}
\end{table}
\end{center}



\subsection{Validated projection of the bipartite network and users polarization}
Because of the official certification released by Twitter about the authenticity of an account, users can be divided into two sets, the verified and unverified ones. In~\cite{Becatti2019}, by implementing the method of~\cite{Saracco2016}, the authors used this feature to infer the accounts' inclination towards a specific political area,  directly from data. The underlying idea is that unverified users follow and interact with verified users sharing their political ideals. In this sense, if two verified users have a high number of common followers and followees, they probably have a similar political affiliation. The \emph{a posteriori} analysis  of the results of the validated projection confirms the previous hypothesis. Due to the Twitter verification procedure, only the information provided by verified users is fact-checkable, thus our check is restricted to this class of users.

We have to pay attention to the contribution of remarkably active users. If a verified user is extremely engaged in the political propaganda, it may interact with a huge number of unverified ones and may thus share a great amount of contacts with almost all other verified users, even those with an opposite political inclination. In this case, the contribution should be considered spurious, being just due to fame of the user. Analogously, the role of an unverified user that retweets all messages from her/his contacts should be discounted.

We obtain the political affiliation of the accounts by considering the undirected bipartite network of interactions (i.e., retweets) between verified and unverified users, aggregated over the whole period: we disregard the information about the direction of the interactions, since we are just interested in groups of users sharing contents. The previous intuition leads to comparing the overlap of connections (literally, the number of common followers and followees) in the real network with the expectations of a null-model able to account for the degree sequence of both layers. In this way, we are able to discount the random noise due to the activity of users and get the significant information from the data. The entropy-based Bipartite Configuration Model (\emph{BiCM},~\cite{Saracco2016}) provides the correct benchmark for this analysis. While we describe more extensively the theoretical construction in the Supplementary Materials, here we outline introduce the main intuitions behind the Bipartite Configuration Model and its monopartite validated projection.

\subsubsection{The Bipartite Configuration Model}
Let us start from a (real) bipartite network and call the two layers L and $\Gamma$ and their dimension respectively $N_\text{L}$ and $N_\Gamma$; we label the nodes on those layers respectively with Latin and Greek indices. We represent the connection via the biadjacency matrix, i.e. the rectangular $(N_\text{L}\times N_\Gamma)$-matrix $\textbf{M}$ whose generic entry $m_{i\alpha}$ is 1 if there is a link connecting node $i\in \text{L}$ and node $\alpha\in\Gamma$, and 0 otherwise. We then consider the \emph{ensemble} $\mathcal{G}_\text{Bi}$ of all possible graphs with the same number of nodes on the two layers as the real network. If we assign a (formal) probability per graph, we can maximize the (Shannon) entropy,
\begin{equation*}
    S=-\sum_{G_\text{Bi}\in\mathcal{G}_\text{Bi}} P(G_\text{Bi})\ln P(G_\text{Bi}),
\end{equation*}
constraining the average value of some quantities of interest on the entire ensemble. If, as it is the case of the present article, we impose the ensemble to have fixed average for the number of links per node (i.e. the degree), the probability per graph factorizes in independent probabilities per link:
\begin{equation}\label{eq:bicm_prob}
    p_{i\alpha}=\dfrac{x_iy_\alpha}{1+x_iy_\alpha},
\end{equation}
where $p_{i\alpha}$ is the probability of finding a link between $i$ and $\alpha$ and $x_i$ and $y_\alpha$ (the \emph{fitnesses}~\cite{Caldarelli2002}) are quantities defined per node that encode the attitude of the nodes to form links~\cite{park2004statistical}. 
At this level, the previous definition is formal, since we just imposed that the average (over the ensemble) of the degree sequence to be fixed, but we did not decide its value. It can be shown (see the Supplementary Materials) that maximizing the likelihood of the real network is equivalent to fixing the average of the degree sequence to the one measured on the real network~\cite{Garlaschelli2008, squartini2011analytical}.

\subsubsection{Monopartite validated projection}
 We can now highlight all contributions that cannot be related to the degree sequence only, comparing the real network with the expectations of the BiCM. Following this line, in~\cite{Saracco2016} a validated projection was proposed on top of the BiCM. The main idea is to consider the common links of two nodes on the same layer and compare it with the theoretical distribution of the BiCM: if the real system shows a commonality of links that cannot be explained only by the activity of the users and we project a link between the nodes under analysis. Using the formalism of~\cite{Saracco2015a},  we call \emph{V-motif}  the overlap.
 
In formulas, by using the independence of probabilities per graph (\ref{eq:bicm_prob}), the probability that both node $i$ and $j$ link the same node $\alpha$ is simply
\begin{equation*}
    p(V^{ij}_\alpha)=p_{i\alpha}p_{j\alpha},
\end{equation*}
where $V^{ij}_\alpha$ is the above mentioned V-motif among $i,\,j$ and $\alpha$. The total overlap between $i$ and $j$ is simply $V^{ij}=\sum_\alpha V_\alpha^{ij}$ and, according to the BiCM, is distributed as a Poisson-binomial, i.e. the extension of a binomial distribution in which all the events have a different probability~\cite{Hong2013}. We can further associate p-values to the observed V-motifs, i.e. the probabilities of finding a number of V-motifs greater or equal to  the one measured on the real network. In order to state the statistical significance of several p-values at the same time, we relied on a multiple test hypothesis and FDR~\cite{benjamini1995controlling} is generally considered the most effective one since it permits to control the number of false negatives, without being too conservative. The result of the projection is a binary undirected monopartite network of nodes from the same layer, that are linked if their similarity cannot be explained only by their degree. We therefore apply the Louvain community detection algorithm~\cite{Blondel2008}  to identify different groups of nodes with similar behaviours in the bipartite network. Since this method is known to be order dependent~\cite{Fortunato2010}, we apply it several times after reshuffling the node order and take the maximum value of the modularity, i.e. the algorithm objective function~\cite{Fortunato2010}.

\subsection{Extraction of the backbone of the traffic activity on Twitter}
In studying the exchange of contents we are interested in the flow that cannot be related by the random activity. Differently from other analogous studies~\cite{Bovet2018,Bovet2019,Stella2018,Stella2018a}, we consider also the virality of tweets, like in~\cite{Becatti2019}. Methodologically, the approach is similar to the one described in the previous section for the extraction of the membership of users, but for substituting the BiCM with its analogous directed version, the Bipartite Directed Configuration Model (\emph{BiDCM},~\cite{DeJeude2018}) and for considering different layers of different kind: while in the section above the layers represent verified and unverified users, here they represent users (both verified and not) and tweets. 
The validated projection procedure returns a \emph{directed} monopartite network of significant exchange of messages, in which the arrow goes from the author to the retweeters. Additional details about this procedure can be found in the Supplementary Materials.

 \section*{Data availability statement} The data that support the findings of this study are available from the corresponding
author upon reasonable request.

\section*{Acknowledgements}
 FDV and GC acknowledge support by EC TENDER SMART (Grant No. 2017/ 0090 LC-00704693). FS and GC acknowledge  the EU project SoBigData (Grant No. 654024). RDN, GC, MP and FS acknowledge support from the Project TOFFEe (TOols for Fighting FakEs) funded by IMT School for Advanced Studies Lucca. All the authors acknowledge enlightening discussions with Andrea Nicolai and Simona De Rosa from the Social Observatory for Disinformation and Social Media Analysis (SOMA) and with Filippo Menczer from the School of
Informatics, Computing \& Engineering of Indiana University.
 
\section*{Declaration of authors}
FDV contributed to the acquisition of data. FDV and FS contributed to the analysis of data. MP and FS contributed to the conception and design of work, and to the manuscript draft. GC, RDN, MP and FS contributed to the interpretation of data and to the manuscript revision. 

Each author approved the submitted version and agreed both to be personally accountable for the author's own contributions and to ensure that questions related to the accuracy or integrity of any part of the work, even ones in which the author was not personally involved, are appropriately investigated, resolved, and the resolution documented in the literature.

\bibliographystyle{naturemag}

\appendix 

\section{Entropy-based null-models}
\subsection{The Bipartite Configuration Model}
In the present section, we  outline the procedure to obtain the probability per link of the Bipartite Configuration Model (\emph{BiCM},~\cite{Saracco2015a}), i.e., the extension of the entropy-based null-models to bipartite networks~\cite{Squartinia,Cimini2018}.

Let us start with a real bipartite network in which we label the two layers with L and $\Gamma$. We will refer to $N_\text{L}$ and $N_\Gamma$, respectively, for the dimensions of the two layers; nodes on the two layers are indicated respectively with $i$ and $\alpha$. A bipartite undirected network can be described in terms of a biadjacency matrix, i.e., a $N_\text{L}\times N_\Gamma$ matrix $\textbf{M}$ whose generic entry $m_{i\alpha}$ is 1 if a link connects the node $i\in\text{L}$ to the node $\alpha\in\Gamma$ and 0 otherwise. The degree of a generic node $i\in\text{L}$, i.e., the number of links of $i$, can be expressed as $k_i=\sum_{\alpha\in \Gamma}m_{i\alpha}$; the analogous for a generic $\alpha\in\Gamma$ is $k_\alpha=\sum_{i\in\text{L}}m_{i\alpha}$.

Let us define $\mathcal{G}_\text{Bi}$ the \emph{ensemble} of all possible graphs with the same number of nodes per layer as in the real network. We label with an asterisk $*$ all the quantities of the real network, but for $N_\text{L}$ and $N_\Gamma$, since they are going to be fixed on the entire ensemble. If every graph $G_\text{Bi}\in\mathcal{G}_\text{Bi}$ is equipped with a probability $P(G_\text{Bi})$, we can define the Shannon entropy for the ensemble as 
\begin{equation*}
    S=-\sum_{G_\text{Bi}\in\mathcal{G}_\text{Bi}}P(G_\text{Bi})\ln P(G_\text{Bi}).
\end{equation*}
$S$ represents the uncertainty we have over the system and we aim at maximising it in order to be as general as possible. However, we discount some information about the system, thus maximising our uncertainty but for some quantities that we want to conserve over the ensemble. This goal can be obtained by maximising the Shannon entropy, constrained such that the average values  of some quantities of interest  are fixed. If the constraints are expressed as a vector $\vec{C}(G_\text{Bi})$ it can be shown that the probability per graph reads~\cite{park2004statistical}:
\begin{equation*}
    P(G_\text{Bi})\sim e^{-\vec{\theta}\cdot\vec{C}(G_\text{Bi})},
\end{equation*}
where $\vec{C}(G_\text{Bi})$ is the vector of the constraints evaluated on the graph $G_\text{Bi}$ and $\vec{\theta}$ is the vector of the relative Lagrangian multipliers~\cite{park2004statistical}. If we constrain the value of the degree sequence, i.e., all $k_i$'s and $k_\alpha$'s, it can be shown that the probability per graph factorises in independent probabilities per link:
\begin{equation*}
    P(G_\text{Bi}|\vec{x}, \vec{y})=\prod_{i\in\text{L}}\prod_{\alpha\in\Gamma}p_{i\alpha}^{a_{i\alpha}}(1-p_{i\alpha})^{1-a_{i\alpha}},
\end{equation*}
where
\begin{equation}\label{eq:p}
    p_{i\alpha}=\dfrac{x_iy_\alpha}{1+x_iy_\alpha}.
\end{equation}
If $\vec{\theta}$ and $\vec{\eta}$ are the Lagrangian multipliers of the degree sequences of the L and $\Gamma$ layers, $x_i=e^{-\theta_i}$ and $y_\alpha=e^{-\eta_\alpha}$. $x_i$ and $y_\alpha$ are called \emph{fitnesses}. Fitnesses encode the attitude, respectively of nodes $i$ and $\alpha$, to establish links~\cite{Caldarelli2002}.

Equation (\ref{eq:p}) is formal: so far, we just impose that, over the ensemble, the average of the degree sequence must be fixed. In order to `tailor' the ensemble on the real network, we have to impose that the average degree sequence is the one measured on the real network~\cite{Garlaschelli2008,squartini2011analytical}, i.e., 
\begin{equation}\label{eq:likelihood}
\left\{
\begin{array}{c}
     \langle k_i\rangle=\sum_\alpha p_{i\alpha}=k_i^*  \\
     \langle k_\alpha\rangle=\sum_i p_{i\alpha}=k_\alpha^* 
\end{array}
\right. .
\end{equation}
In this way, we have an ensemble of networks to compare our results with, showing exactly the same degree sequence, but being as general as possible regarding all the other features. 

The solution of System (\ref{eq:likelihood}) can only be numerical, since no analytic form can be found. Due to the  dimensions of the system under analysis, we approximate the probability per link as $p_{i\alpha}\simeq x_iy_\alpha$; the approximation is justified as long as the network is sparse, in particular when $x_i, y_\alpha\ll1$. In such a case, the null model is analogous to the one of~\cite{Chung2002} and the solution of System (\ref{eq:likelihood}) is analytic: 
\begin{equation}\label{eq:CLA}
    p_{i\alpha}^\text{CLA}=\dfrac{k_i^*k_\alpha^*}{m^*},
\end{equation}
where $m^*$ is the total number of links measured on the real network and ``CLA" stays for ``Chung-Lu Approximation".

\subsubsection{Validated monopartite projection}
We can make use of the {\it BiCM} in order to project the information contained in the original network, discounted by the degree sequence contribution. The main idea was developed in~\cite{Saracco2016}: due to the independence of the probabilities per link, the probability that both nodes $i,j\in\text{L}$ are connected to node $\alpha\in\Gamma$ is
\begin{equation*}
    P(V^{ij}_\alpha)=p_{i\alpha}p_{j\alpha},
\end{equation*}
where $V^{ij}_\alpha$ is the event of $i$ and $j$ linking $\alpha$: such a pattern was described in~\cite{Saracco2015a} as a $V$-motif since, if the two layers are represented as two horizontal lines of vertices, it draws a ``V" between the layers. More formally, they are $K_{2,1}$ stars~\cite{Diestel2012}. If we are interested in how much similar two nodes are in terms of common connections, we can compare the number of common links in the real network $(V^{ij})^*=\sum_{\alpha\in\Gamma}(V^{ij}_\alpha)^*$ with its theoretical distribution, according to the {\it BiCM}. Since all the probabilities $P(V^{ij}_\alpha)$ are different -in general- the theoretical probability distribution is a Poisson-binomial~\cite{Volkova2005, Deheuvels1989a, Hong2013}, i.e., the extension of a Binomial distribution in which every Bernoulli event has a different probability. If the probabilities per $V-$motif are particularly low (as it is the case of the present study), the Poisson-binomial distribution can be approximated with a Poisson distribution~\cite{Hong2013}, in which the parameter is
\begin{equation*}
    \lambda=\langle V^{ij}\rangle=\sum_{\alpha\in\Gamma}p_{i\alpha}p_{j\alpha}\overset{\text{CLA}}{=}\dfrac{k_i^*k_j^*}{(m^*)^2}\sum_{\alpha\in\Gamma}(k_\alpha^*)^2,
\end{equation*}
where in the last step we implement the Chung-Lu approximation (\ref{eq:CLA}).

Once we have all the theoretical distributions, we can assign to every $V^{ij}$  a p-value and then validate all the statistically significant ones. In the present article, we set the statistical significant level $\alpha=0.01$ and implement the FDR (\emph{False Discovery Rate}, \cite{benjamini1995controlling}) for the validation. FDR is one of the most effective methods for multiple testing hypothesis, i.e., to claim the statistical significance of many p-values at the same time, since it permits to control the number of False Positives~\cite{benjamini1995controlling}. The procedure is pretty simple: once all N p-values are ordered from the smallest to the greatest, i.e.,
\begin{equation*}
    \text{p-value}_1\le\text{p-value}_2\le\dots\le\text{p-value}_N,
\end{equation*}
the effective statistical significance threshold is the greatest $i\frac{\alpha}{N}$ satisfying
\begin{equation*}
    \text{p-value}_i\le i\dfrac{\alpha}{N}.
\end{equation*}

\subsection{The Bipartite Directed Configuration Model}
As a side product of defining a  monopartite Configuration Model with a community structure, in~\cite{DeJeude2018} the directed extension of the {\it BiCM} was presented; hereafter, we are going to call it Bipartite Directed Configuration Model ({\it BiDCM}). In the following, we will use the formalism of~\cite{Becatti2019}, since it is tailored on our application. Thus, we will indicate the two layers as $U$ (i.e., users) and $P$ (i.e., posts), and $N_U$ and $N_P$ are, respectively, their dimensions. A bipartite directed network can be described by using two $N_U\times N_P$ biadjacency matrices, respectively $\textbf{T}$ (i.e., the tweets), indicating links going from $U$ to $P$, and $\textbf{R}$ (i.e., retweets), indicating links from $P$ to $U$. Considering the exchange of content in the original network, we use the $\textbf{T}$ matrix to describe the authorship of tweets, while $\textbf{R}$ is for retweets. If we focus on a generic user $u$, her/his out- and in- degrees are respectively
\begin{equation*}
    \kappa_u^\text{out}=\sum_{p\in P}t_{up};\qquad\kappa_u^\text{in}=\sum_{p\in P}r_{up}
\end{equation*}
The two quantities represent, respectively, the number of tweets the user $u$ wrote and the number of tweets $u$ retwitted.
Analogously, for the generic post $p$ is
\begin{equation*}
    \kappa_p^\text{in}=\sum_{u\in U}t_{up}=1;\qquad\kappa_p^\text{out}=\sum_{u\in U}r_{up}.
\end{equation*}
As it can be seen in the previous definitions, we have a great simplification, due to the nature of our system: since there can only be a single author for a given post, the in-degree of a post is always 1, for every $p$. This feature is going to introduce a great simplification in the following calculations.

By following the same track of the previous section, if we call the ensemble of bipartite directed graphs $\mathcal{G}_\text{BiD}$, by maximising the entropy of the system constraining the entire (directed) degree sequence, it can be shown that the probability for a generic bipartite directed graph $G_\text{BiD}\in\mathcal{G}_\text{BiD}$ still factorises in terms of probabilities per link~\cite{DeJeude2018,Becatti2019}:
\begin{equation*}
    P(G_\text{BiD}|\vec{z}, \vec{z}', \vec{\zeta}, \vec{\zeta}')=\prod_{u\in U}\prod_{p\in P}q_{up}^{t_{up}}(1-q_{up})^{1-t_{up}}\cdot\prod_{u'\in U}\prod_{p'\in P}(q_{u'p'})^{r_{u'p'}}(1-q_{u'p'})^{1-r_{u'p'}},
\end{equation*}
where
\begin{equation}
    q_{up}=\dfrac{z_u\zeta_p}{1+z_u\zeta_p};\qquad q'_{u'p'}=\dfrac{z'_{u'}\zeta'_{p'}}{1+z'_{u'}\zeta'_{p'}}
\end{equation}
are respectively the probabilities of $u$ being an author of the post $p$ and of $u'$ being a retwitter of the post $p'$. Again, by maximising the likelihood of the real systems, we have that, among the other conditions, $\langle \kappa_p^\text{in}\rangle=1\,\forall p\in P$. The previous condition implies that all the fitnesses $\zeta$ are equal for every node in $P$ and it can be shown that $q_{up}=\frac{(\kappa_u^\text{out})^*}{N_P}\,\forall p\in P$~\cite{Becatti2019}.

\subsubsection{Directed validated projection}
 In order to extract those relationships among users that cannot be simply referred to their activity and the virality of posts, we project the directed bipartite network on the layer of users, following the same strategy of the validated projection for undirected bipartite networks. Differently from the previous validated projection, in the present case the final result is going to be a directed monopartite network in which links flow from the significant source of post to the significant retwitters.

In the present study, the approach is analogous to the one in~\cite{Becatti2019}, but for two crucial differences. First, in~\cite{Becatti2019} we analysed the contribution of retwitted posts only, while here we also consider the contribution of non-retwitted ones. This choice affects only the probabilities $q$'s: the rationale is that we aim at detecting if an account retweets all the posts of a certain user or it makes a selection on the messages of the given author. Secondly, we not only limit the investigation to the content flow, but also we explore the contribution of bot accounts to that flow. In terms of message spreading, and linked to our exploration on the activities, retwitting all the posts of a certain user is probably more critical than selecting part of them.


We call $\mathcal{V}^{uv}_p$ the event of $v$ retwitting the post $p$ written by $u$: according to the {\it BiDCM}, its probability is simply
\begin{equation}\label{eq:directed_v}
    P(\mathcal{V}^{uv}_p)=q_{up}q_{vp}'=\dfrac{(\kappa_u^\text{out})^*}{N_P}q_{vp}'
\end{equation}
As in the case of the undirected bipartite network, due to the dimension of the system, we approximate the Poisson-binomial distribution of $\mathcal{V}^{uv}$ with a Poisson distribution in which the parameter is \begin{equation*}
    \lambda=\langle\mathcal{V}^{uv}\rangle=
    \sum_{p\in P}P(\mathcal{V}^{uv}_p)=
    \sum_{p\in P}\dfrac{(\kappa_u^\text{out})^*}{N_P}q_{vp}'=
    \dfrac{(\kappa_u^\text{out})^*}{N_P}\sum_{p\in P}q_{vp}'=
    \dfrac{(\kappa_u^\text{out})^*(\kappa_v^\text{in})^*}{N_P},
\end{equation*}
where, in the last steps, we made use of equation (\ref{eq:directed_v}). It is worth noting that, with respect to the previous case, in the actual projected validation we do not use the Chung-Lu approximation in order to simplify the calculations, but we take advantage of the fact that any post has just one author.

Finally, we can associate to each observed $\mathcal{V}^{uv}$ a p-value via the previous distributions and we validate them via the above mentioned FDR~\cite{benjamini1995controlling}; in the present case too, the statistical significant level is set to $\alpha=0.01$.

\section{Iterative procedure for assigning a political inclination to unverified users}
As mentioned in the main text, just almost one half of the unverified users interact with verified ones. Since this interaction is used to infer the inclination of users towards a political point of view
directly from data, we iteratively extend the procedure for the inclination assignment. The idea is to consider all the unpolarized users and assign them the inclination of the majority of their followers and followees. If the polarization index $\rho$ (as defined in (1) of the main text) of a given unpolarized node is lower than 0.5, then the majority of its neighbours is non polarized. Thus, the node re-enters the polarization step, until no polarization can be assigned anymore.  Noticeably, the previous threshold -0.5- is particularly high, since we have identified 3 big communities. The nodes resulting in grey in the first bipartite validation have been considered in the contagion polarization too.

The algorithm stops after 10 steps, and we were able to assign a political affiliation to almost the 27\% of the unpolarized users, resulting in the 15\% of the entire set of accounts. Even if the previous percentage seems quite small, it is not when we focus on the nodes in the directed validated projection, as it can be seen in Figure 5 of the main text. The amount of unpolarized users represents the 8.3\% of the total number of validated accounts. Literally, the polarization by contagion was able to assign a polarization to almost the 58\% of unpolarized nodes in the directed projection.

\section{Alternative approaches to assign a political inclination}
\begin{figure}[ht!]
    \centering
    \begin{minipage}{.48\textwidth}
    \centering
    \includegraphics[width=\linewidth]{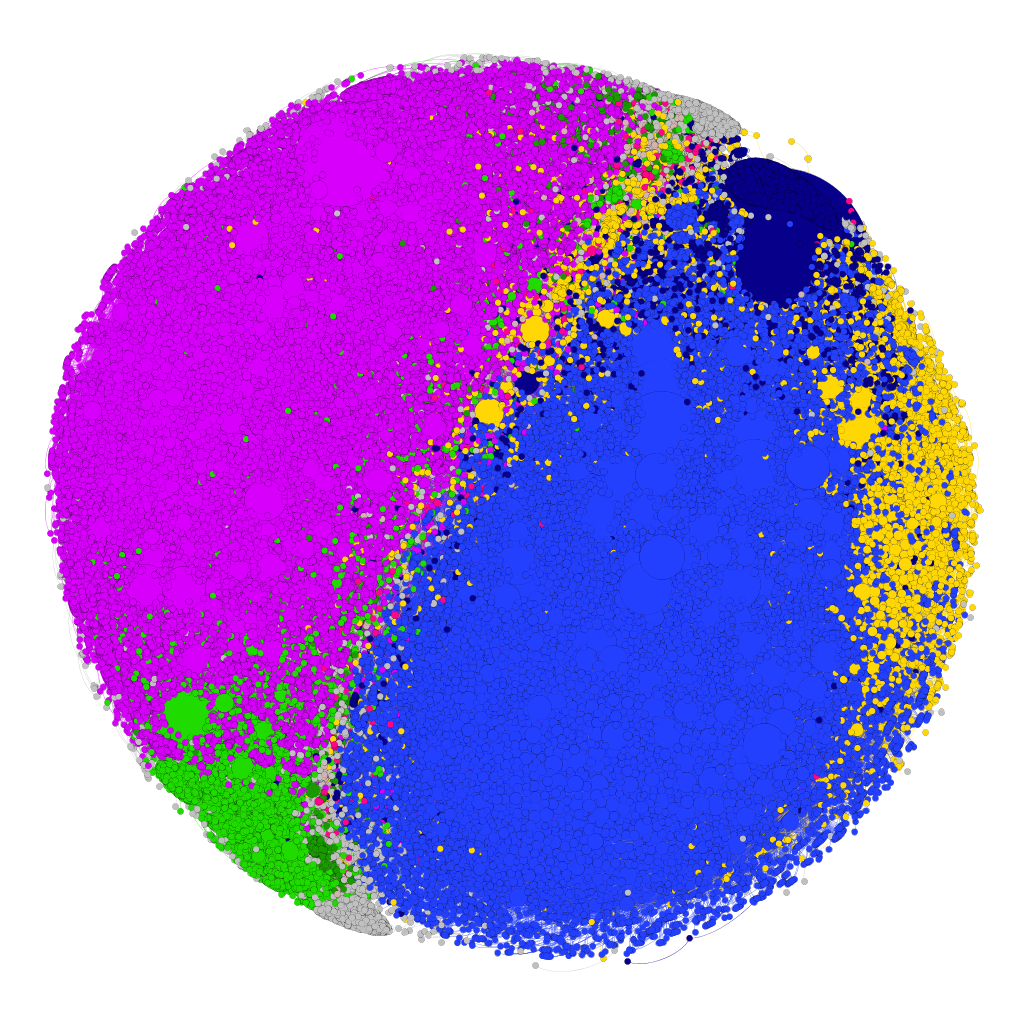}
    \end{minipage}
    \hfill
    \begin{minipage}{.48\textwidth}
    \centering
    \includegraphics[width=1\linewidth]{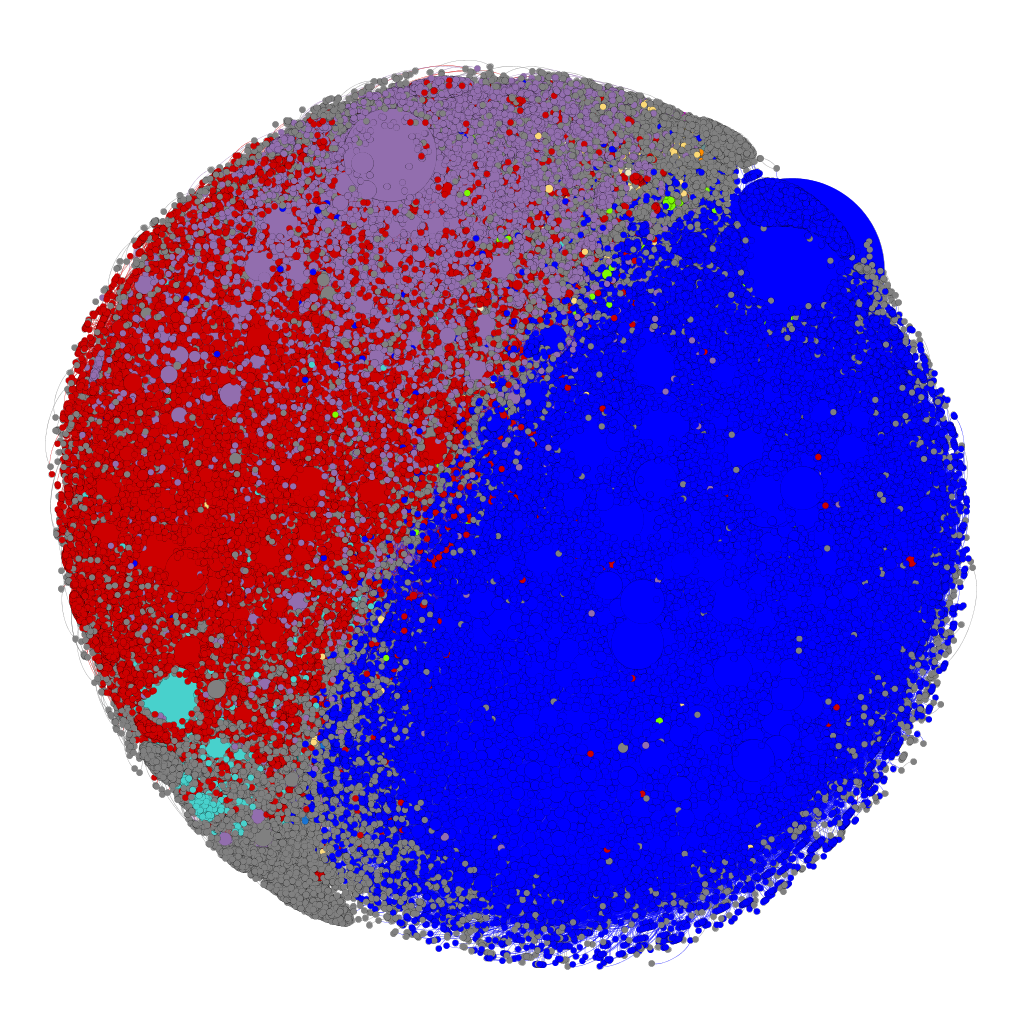}
    \end{minipage}
    \caption{The undirected network of retweet interactions. Left panel: Nodes are colored according to the reshuffled community detection on the network. Right panel: Nodes are colored according to the iterative polarization assignment described in the main text. }
    \label{fig:menczener}
\end{figure}

The adoption of the validated projection approach for detecting the polarization of the users may appear as unnecessarily complicated, with respect to launching a community detection algorithm on the monopartite network of interactions among all the accounts. However, the validation process removes random noise from the system, thus providing a (validated) network in which the communities are clearer. The monopartite network of undirected interactions can be found in Figure~\ref{fig:menczener}: in the left panel, nodes have been colored according to the result of the Louvain community detection  algorithm;  in the right panel, the communities have been obtained by the (iterative) polarization described in the main text. There are some similarities between the two partitions: the blue communities almost overlap (but for some relevant differences that we are going to discuss in the following). However, on the one hand, the left partition pinpoints a yellow group (containing mostly M5S politicians and journalists); on the other hand, the right partition reveals two main groups in the upper cluster, instead of the single one on the left.

At first glance, the two panels may seem more or less equivalent, but literally they are not. If we consider the affiliation of the various verified users, the differences are somehow striking. Firstly, the pink community in the left panel contains the verified accounts of both the red and the purple community of the right panel, thus mixing the two groups. Secondly, the yellow community in the left partition reveals a group that is not detected in the right one, i.e., the one of politicians and journalists supporting the Movimento 5 Stelle (M5S) party. However, this cluster also contains other newspapers 
far from M5S, as `La Stampa', `Il Corriere della Sera', `Il Sole24ore', a gossip website (`Dagospia') and some TV newscasts as `SkyNews 24', `RaiNews' and `Tg La7'. The blue community in the left panel is quite similar to the one in the right panel, but for the presence of some Democratic Party parliamentarians and for the absence of Mr. Matteo Salvini. The latter fact is especially odd, since the two official accounts of Lega, i.e., the political party of Mr. Salvini, are still in the blue community. In the dark blue community of the left panel, absent in the right panel, we find the account of Mr. Salvini, together with some French politicians from the \emph{Rassemblement National}, as well as some German politicians from the \emph{Alternative f\"ur Deutschland}. At a first sight, this sounds reliable, since Lega, Rassemblement National, Alternative f\"ur Deutschland are allied in the European Parliament. Nevertheless, the presence of some parliamentarians of the Italian Democratic Party  weaken the reliability of such a cluster. Analogously, the quite big green community in the left panel contains a pot-pourri of accounts, like, e.g., the Italian embassy in Greece and the referees of a popular Italian TV dance contest, `Ballando con le Stelle'.\\

We argue that the non validated partition, e.g., the one described in the left panel, is reasonable, but the amount of noise is much higher with respect to the validated one, i.e., a relatively high number of verified nodes has a clearly wrong membership. In the validated projection, the number of evidently wrong assignments is much lower, being limited to few center-left wing journalists in the community of the right wing. Due to such a difference, in this paper we have adopted the validated projection approach.

\section{Bot Squads}

The analysis of the role of bots in the validated backbone of the traffic of messages on Twitter let us observe the presence of automated accounts that follow more than one genuine user. We call such groups `bot squads'. Remarkably, we find two groups of genuine users sharing more than 2 automated accounts. The biggest users group includes 12 of the top 20 hubs (9 in the top 10). 
Referring to Figure 7 in the main text, we detect the presence of a subgraph of 22 hubs sharing 22 bots among their followers: literally, each of these 22 bots does not retweet  the content of a single hub, but of more than one user in the subgraph.  
 Thus, the bot squad increases the visibility of the common followees.
%
Among the hubs in the subgraph, we can find the account of the Minister of Internal Affairs Mr. Salvini, as well as the Minister of Infrastructures Danilo Toninelli. Other verified accounts are the one of Giorgia Meloni of Fratelli d'Italia, a right wing party, and the official accounts of Lega (Mr. Salvini's party). In this set, we find also the news website supported by Casa Pound, a neo-fascist Italian party, two journalists from the same website, the director of the newspaper La Verit\`a, as well as other politicians of the Lega party. 

The Figure 8 in the main text shows the subgraph of the violet bot squad: there, the genuine accounts are much less efficient in delivering their messages, since the strongest hub, representing a popular newscast, ranks $176^\text{th}$ in the overall hub score. Moreover, while the blue subgraph  included 178 nodes, the violet one contains just 58 accounts.  

In this subgraph, we find the presence of several NGOs, some NGO representatives (coming, for instance, from `Comunit\`a di Sant'Egidio', a Catholic NGO, and from MOAS, a NGO active in aiding migrants in the Mediterranean sea), some journalists from the TV channel La7 and small political parties belonging to the left wing.

The incidence of bots in the two subgraphs of Figures 7 and 8 in the main text is relatively similar, being the 87\% for the first subgraph and 79\% for the second one. Instead, the ratio between the members of the bot squads (i.e., the number of shared bots) over the total number of genuine users in the subgraphs is not: the former is exactly 1, while the latter is around 0.58. Interestingly, in both sets, the hubs rarely retweet between each other in a significant way (in fact, only 3 links can be found among them). They leave the duty of spreading the content of the partners to the bots.

\end{document}